\documentclass[11pt]{article}
\usepackage[margin=1in]{geometry}
\usepackage{amsmath,amssymb}
\usepackage[utf8]{inputenc}
\usepackage[T1]{fontenc}
\usepackage{natbib}
\usepackage{graphicx}
\usepackage{booktabs}
\usepackage{threeparttable}
\usepackage{array}
\usepackage{tabularx}
\usepackage[table]{xcolor}
\usepackage{lscape}
\usepackage[figuresright]{rotating}
\usepackage{hyperref}
\hypersetup{hidelinks}
\providecommand{\pagerange}[1]{}\providecommand{\pubyear}[1]{}
\providecommand{\volume}[1]{}\providecommand{\artmonth}[1]{}\providecommand{\doi}[1]{}
\providecommand{\backmatter}{}
\providecommand{\tightlist}{\setlength{\itemsep}{0pt}\setlength{\parskip}{0pt}}
\newenvironment{keywords}{\par\medskip\noindent\textbf{Key words:} }{\par\medskip}

\definecolor{cfBlue}{HTML}{0072B2}
\definecolor{warnRed}{HTML}{D55E00}
\definecolor{headGrey}{HTML}{E8ECF0}
\definecolor{cfTint}{HTML}{D6E4F0}
\definecolor{warnTint}{HTML}{F6D9C9}

\newcommand{\fitwidth}[1]{\resizebox{\ifdim\width>\linewidth\linewidth\else\width\fi}{!}{#1}}

\title{Best for which estimand? A known-truth benchmark of
longitudinal-matching and target-trial-emulation methods for time-varying treatments}

\author{M. Ehsan Karim\\[4pt]
\normalsize
\begin{minipage}{0.86\textwidth}\centering
School of Population and Public Health, University of British Columbia, Vancouver, British Columbia, Canada;
and Centre for Advancing Health Outcomes, St.\ Paul's Hospital, Vancouver, British Columbia, Canada\\[3pt]
\texttt{ehsan.karim@ubc.ca}\\[2pt]
\href{https://orcid.org/0000-0002-0346-2871}{ORCID: 0000-0002-0346-2871}
\end{minipage}}
\date{}

\begin{document}
\maketitle

\begin{abstract}
On a non-collapsible survival mechanism, longitudinal-matching and target-trial-emulation methods are not
competing estimators of one truth but answers to \emph{different causal questions}, so a benchmark that scores
them against a single ``true hazard ratio'' fabricates bias. We provide the direct-comparison benchmark of relative efficiency, variance estimation, and model sensitivity that recent methodological reviews find lacking. On a deliberately non-collapsible continuous-time Cox
data-generating mechanism with known truth, we show that the dominant families (sequential Cox, sequential
stratification, risk-set matching, and inverse-probability-of-treatment-weighted (IPTW) marginal structural models) target \emph{numerically distinct} causal
estimands (marginal, conditional, two average-treatment-effect-on-the-treated, and intention-to-treat versus
per-protocol). First,
we quantify the \emph{phantom bias} a single shared marginal truth fabricates: $0.32$--$0.33$
log-cumulative-hazard-ratio units for the matching estimators and $0.15$ for the conditional method; the
associational naive time-dependent Cox sits $0.76$ away, a total discrepancy compounding the
estimand gap with confounding. Second, a \emph{rank reversal}: the recommended method flips with the target estimand, and a low-variance
off-target estimator can still win on mean-squared error. Third, a \emph{cross-family variance result}: in our simulations the cluster-robust sandwich is closer to nominal for
the trial-stacking estimator ($0.90$) but under-covers the matching estimators ($0.77$--$0.82$), which a
prespecified $n{=}500$ bootstrap sub-study brings to $0.95$--$0.96$. Fourth, \emph{model sensitivity}: omitting a confounder induces $0.45$--$0.50$ log-hazard-ratio bias and
undercoverage, and intention-to-treat and per-protocol effects diverge as switching increases; a Stanford
heart-transplant analysis illustrates these. On a second mechanism three of the four findings replicate,
the rank reversal attenuating and model sensitivity proving calibration-dependent.
\end{abstract}


\section{Introduction}\label{sec:intro}

\textbf{The question, and the reframe.} Faced with a dozen ways to analyse a time-dependent treatment, the
practical question is \emph{which method is best?} We argue---and show on a mechanism with known truth---that this
is the wrong question. On a non-collapsible survival model the leading methods are not competing estimators of one
number but consistent estimators of \emph{different causal estimands}, so the well-posed question is \emph{best for
which estimand?} Apparent bias and method rankings turn out to depend on which estimand was asked for; interval coverage depends additionally on how each construction's sampling variability is estimated. This paper makes that dependence
quantitative on a deliberately non-collapsible Cox mechanism where every estimand's truth is known.

\textbf{Background and motivation.} When a treatment is initiated at different times over longitudinal follow-up,
a naive comparison of ever- versus never-treated patients is subject to immortal-time bias, and simply entering
treatment as a time-dependent covariate in a Cox model is invalid when time-dependent confounders are themselves
affected by prior treatment \citep{hernan2016,keogh2023}. A large and scattered literature addresses this by \emph{emulating a sequence of
trials}: at each time origin, eligible untreated patients are partitioned into those who initiate treatment
(``treated'') and those who do not (``controls''), followed forward, and combined across origins.
\citet{thomas2020} unify this literature into a class of longitudinal-matching methods whose roughly twenty named
variants collapse, by their own account, into three estimation templates that we carry throughout under the names
of Table~\ref{tab:estimands}: \emph{sequential Cox / target-trial emulation}
\citep{gran2010,hernan2016,danaei2013,keogh2023}, \emph{sequential stratification}
\citep{schaubel2006,schaubel2009,taylor2013,kennedy2010}, and \emph{time-dependent propensity-score
(risk-set) matching} \citep{lu2005,li2001,smith2015}. They close with a comparison gap and three named simulation gaps, verbatim: ``few direct comparisons
exist. Simulation studies comparing the relative efficiency, variance estimation, and model sensitivity are
lacking.''

\textbf{The pitfall we avoid.} A naive head-to-head (run every method on one dataset and compare to a single
true hazard ratio) is uninformative, because on a non-collapsible survival model these methods \emph{target
different causal estimands}, so a shared truth manufactures bias. Conversely, scoring each method against
\emph{its own} probability limit is a tautology: a consistent, correctly coded estimator cannot fail that gate.
We therefore treat the own-limit comparison purely as a \emph{correctness gate} and locate the paper's
contribution elsewhere.

\textbf{Contributions.} On a non-collapsible Cox mechanism we deliver four empirical findings.
\begin{enumerate}\tightlist
\item \textbf{Phantom-bias magnitude.} The decision-relevant bias a single-shared-``true-HR''
comparison fabricates when an average-treatment-effect-on-the-treated (ATT) or conditional method is scored
against a marginal truth.
\item \textbf{Rank reversal.} The recommended method flips with the target
estimand; we report both the estimand-aligned winner (whose probability limit is closest to the requested
estimand) and a mean-squared-error (MSE) winner, and we show that a precise but
differently-targeted estimator can beat a noisier on-target one on MSE, the phantom-bias logic applied to method
choice. This within-class comparison is one that recent work on nested-versus-single-point estimands
\citep{wiener2026} does not provide.
\item \textbf{Cross-family variance result.} The cluster-robust
sandwich covers the trial-stacking estimator but under-covers the matching estimators, for which the
subject-level bootstrap restored near-nominal coverage in the cells we examined, analogous to the matching-uncertainty problem of \citet{abadie2006} (an empirical repair whose general validity we do not establish; Web Appendix~B).
\item \textbf{Model-sensitivity and intention-to-treat-versus-per-protocol} arms that quantify two axes
the estimand view does not by itself capture.
\end{enumerate}
We provide the estimators as the R package \texttt{lmtte},
including new implementations of sequential stratification and risk-set caliper matching.

\textbf{Positioning.} Our closest neighbours each miss a different axis. \citet{wiener2026} compare
sequential-nested versus single-point trial estimands under effect modification but do not compare the method
families against one another. \citet{limozin2025} develop inference procedures for sequential trial emulation with
survival outcomes, but for a single family rather than across them, and without a known-truth benchmark that
separates estimand gaps from estimator error. \citet{richey2024} compare time-varying
propensity-score matching to sequential stratification, but for a \emph{continuous} outcome with no effect
modification and only two method families. \citet{thomas2020} is a qualitative review without known truth. Our companion
work treats confidence-interval estimation for one additive-hazards estimator; here the outcome models are Cox /
stratified-Cox throughout and variance is handled with standard estimators, so the two are disjoint.

\textbf{Roadmap.} Section~\ref{sec:estimands} states the estimand each family targets (Table~\ref{tab:estimands})
and Section~\ref{sec:theory} characterizes the estimand structure and its consequences.
Section~\ref{sec:dgm} gives the non-collapsible Cox data-generating mechanism and the correctness gate.
Section~\ref{sec:sim} lays out the ADEMP simulation and its roadmap (Table~\ref{tab:simmap}).
Section~\ref{sec:results} reports the four findings. Section~\ref{sec:applied} illustrates on the Stanford heart
transplant data and Section~\ref{sec:discussion} discusses. Derivations, the pooled simulation results and robustness suite (with the full per-cell values in the code repository; see Data Availability) are in the Supplementary Material (Web Appendices~A--F).

\section{Methods and their estimands}\label{sec:estimands}

\textbf{Setup.} Let $A_k \in \{0,1\}$ be a monotone (once-treated, always-treated) treatment recorded at visits
$k=0,\dots,K-1$, $L_k$ a time-dependent confounder affected by prior treatment, and $V$ a baseline
effect-modifier. Each emulated trial is indexed by its \emph{origin} $s\in\{0,\dots,K-1\}$, the visit at which the
trial starts; $s{=}0$ denotes the origin-$0$ always-versus-never reference contrast (strategies assigned and
followed from the first visit). The methods differ in construction and, crucially, in target estimand; the constructions and the
two variance methods (cluster-robust sandwich and subject-level bootstrap) are detailed in Web Appendix~B. Three routes make the estimands diverge: (1) because the hazard ratio
is non-collapsible \citep{hernan2010}, a covariate-conditional and a marginal hazard contrast differ even absent
confounding; (2) matching targets an effect on the treated; and (3) per-protocol censoring of control initiators changes
the estimand relative to intention-to-treat. Table~\ref{tab:estimands} records, for each method, the target
parameter, the population it is defined over, and the probability limit against which it is scored. No target is
read off a data-generating coefficient; each is computed by replaying the estimator's own construction on a large
pseudo-population (Web Appendix~A).

\begin{table}[htbp]\centering\small
\caption{\textbf{Method families and the causal estimand each targets.} The \emph{Method} and \emph{Specification}
columns name each estimator; we use these names throughout the paper (the three families collapse the roughly
twenty variants reviewed by \citet{thomas2020}). For each we give the target parameter, the population over which
it is defined, and the ``truth'' it is scored against: its own probability limit, obtained by replaying the
estimator's construction on a large pseudo-population, not by reading a data-generating coefficient. ATT, average
treatment effect on the treated; RMST, restricted mean survival time; PP, per protocol; IPCW,
inverse-probability-of-censoring weighting; ITT, intention to treat; TD-PS, time-dependent propensity-score; log-CHR, log-cumulative-hazard ratio. Own-limit scoring is a correctness gate, not a
finding: a consistent estimator cannot fail it. The numerically distinct limits are what make a single ``true
hazard ratio'' comparison misleading. For the standardized Sequential Cox, ITT Sequential Cox, and IPTW-MSM rows, the same estimand is additionally summarized on a fixed-horizon log-cumulative-hazard-ratio (log-CHR) scale, the phantom-bias reference scale and a rank-reversal target used in the results.}\label{tab:estimands}
{\setlength{\tabcolsep}{4pt}%
\begin{tabular}{@{}>{\raggedright\arraybackslash}p{2.3cm}>{\raggedright\arraybackslash}p{2.5cm}>{\raggedright\arraybackslash}p{3.6cm}>{\raggedright\arraybackslash}p{2.4cm}>{\raggedright\arraybackslash}p{2.7cm}@{}}
\toprule
\rowcolor{headGrey} Method & Specification & Target parameter & Population & Truth (own probability limit) \\
\midrule
Sequential Cox & standardized (PP+IPCW) & marginal origin-pooled RMST / survival difference & full trial-eligible, origin-pooled & origin-pooled g-computation replay \\
\addlinespace
Sequential Cox & conditional (PP+IPCW) & conditional log-HR & trial-eligible, baseline-adjusted & plim of the adjusted Cox \\
\addlinespace
Sequential Cox & ITT & marginal RMST of the baseline-assigned strategy & all trial-eligible, origin-pooled & ITT-estimator replay (origin-pooled) \\
\addlinespace
Sequential stratification & PP / as-treated & ATT within-stratum log-HR & treated initiators + risk-set controls & plim of the stratified Cox \\
\addlinespace
TD-PS matching & risk-set caliper & ATT within-set log-HR & matched treated + caliper controls & plim of the matching estimator \\
\addlinespace
IPTW marginal structural model & full-cohort weighting & marginal RMST / survival difference from $s{=}0$ & full cohort from $s{=}0$ & full-cohort always-vs-never pseudo-RCT \\
\addlinespace
Naive time-dependent Cox & reference (non-causal) & association log-HR & full cohort & associational limit (no causal estimand; excluded from the Corollary) \\
\bottomrule
\end{tabular}}
\end{table}

\textbf{Two distinctions are essential for interpreting the comparisons.}
\begin{enumerate}\tightlist
\item Sequential stratification censors a
control at its own initiation and is therefore \emph{per-protocol / as-treated}, not intention-to-treat;
mislabelling it injects a spurious ITT-versus-PP gap.
\item A stratified Cox on matched sets emits a
\emph{conditional} log-HR, so sequential stratification versus TD-PS matching is a construction/weighting contrast on a shared conditional
scale, not a scale change.
\end{enumerate}
We make both precise in Web Appendix~A, where each estimand is defined as the
probability limit of the estimator's own construction.

\section{Estimand structure and its consequences}\label{sec:theory}

\textbf{The differences are structural, not incidental.} The four families do not estimate one quantity with
different efficiency; they target \emph{different functionals of the counterfactual survival curves}. We state that
structure here as a characterization---grounded in known results and confirmed mechanism-by-mechanism on the
known-truth simulation---with the formal statements, assumptions, and numerical confirmation in Web Appendix~A.

\emph{Estimands as functionals.} Writing $S^a(\tau)$ for survival under the static regime $a$ (and labelling the
always-versus-never marginal reference $s{=}0$, i.e.\ the $a{=}1$-versus-$a{=}0$ contrast), the marginal
log-cumulative-hazard-ratio, the covariate-conditional log-HR, the two average-treatment-effect-on-the-treated
(ATT) parameters, and the intention-to-treat (ITT) effect are \emph{distinct estimands}. Only the marginal
reference is a functional of the static-regime curves $\{S^a\}$, while the conditional, ATT and ITT limits are
functionals of the richer counterfactual law (of assignment, adherence, and outcome); no single ``true
hazard ratio'' exists for all of them. Their \emph{causal} reading requires the standard identifying
assumptions: consistency, sequential exchangeability given the measured past, positivity, and no interference / a single version of treatment (the stable-unit-treatment-value assumption, SUTVA)---and, for the matching constructions, an asymptotic within-set balance condition (R4) that pins their ATT limits, collected in Web Appendix~A; these hold by construction throughout the known-truth simulation, including the unmeasured-frailty slice, whose
$U$ is an unmeasured \emph{prognostic} factor balanced across treatment, not a confounder.

\emph{The correctness gate (Web Appendix~A, Lemma~1).} Each correctly-constructed estimator is consistent for its own
probability limit: the trial-stacking estimators by standard $Z$-estimation, the matching estimands defined as the
estimator's limit. Scoring an estimator against its own limit is therefore a correctness gate a consistent
estimator cannot fail, not a finding.

\emph{Estimand separation (Web Appendix~A, Proposition~1).} On the non-collapsible mechanism the estimand set is
\emph{non-degenerate}, separated by four \emph{sufficient, co-occurring} mechanisms: non-collapsibility (a known
survival fact we invoke rather than re-derive), treatment-affected-confounder mediation, effect modification with
covariate selection, and non-adherence. They are illustrative and can act together, not an orthogonal
one-parameter-each decomposition.

\emph{Phantom bias is the estimand gap (Web Appendix~A, Corollary).} It follows that a consistent estimator scored
against a \emph{foreign} estimand's truth is biased by exactly the estimand gap, a \emph{phantom bias} that a
shared-``true-HR'' benchmark misattributes to the estimator. Its magnitude is our first result (\S\ref{sec:results}).

\emph{Rank reversal (Web Appendix~A, Theorem~1).} No estimator minimizes bias for every target: the bias-minimizing
choice is the construction whose limit is nearest the requested estimand, and that choice changes across targets.
Separately, by the bias--variance identity a precise but off-target estimator can win on mean-squared error when the
on-target one is much noisier: the phantom-bias logic applied to method choice, and the reason we report the
bias-aligned and MSE rankings separately (\S\ref{sec:results}).

\emph{Variance (Web Appendix~B, a heuristic).} Finally, the cluster-robust sandwich conditions on the
\emph{estimated} matched sets and so omits their estimation uncertainty, under-covering the matching estimators;
the subject-level bootstrap, which re-forms the sets, repairs it (\S\ref{sec:results}). We treat this as a
heuristic; the influence-function analysis of matching uncertainty is developed in companion work.

\section{Data-generating mechanism}\label{sec:dgm}

\textbf{One fixed non-collapsible population, truth computed once.} We simulate a continuous-time multiplicative
hazard
\[
\lambda(t\mid A_t,L_t,V,U) = \lambda_0 \exp(\beta_A A_t + \beta_{AV} A_t V + \beta_L L_t + \beta_U U),
\]
here $V$ is a baseline effect-modifier and $U$ an unmeasured frailty, and event times are drawn by inverse-CDF
sampling within each unit visit interval, so the conditional Cox model is
exactly correctly specified. The confounder evolves as $L_k = 0.8\,L_{k-1} - \psi A_{k-1} + \text{drift}\cdot k +
\varepsilon_k$; treatment is once-on with initiation probability $\mathrm{expit}(\gamma_0 + \gamma_L L_k +
\gamma_V V)$; administrative censoring occurs at $\tau=K$. The structural effect $\beta_A$ is pinned substantively
nonzero (non-collapsibility vanishes at the null), and the base arm sets $\beta_U=0$ so the Cox limit adjusting for the time-updated confounder $L_t$ equals the structural coefficient. This mechanism is \emph{non-collapsible} by construction: the
marginal and conditional hazard contrasts differ by a known amount even at zero confounding, which is precisely
the setting in which the estimand decomposition can be stated; a multiplicative hazard is always positive, so,
unlike additive-hazard mechanisms, there is no non-negativity/truncation problem. The lineage is the Cox
g-formula simulation of \citet{havercroft2012} and \citet{youngtchetgen2014}, with a treatment-affected
confounder in the spirit of \citet{keogh2023}. The study values are $\lambda_0=0.12$, $\beta_A=-0.70$, $\beta_L=0.5$, $\psi=0.5$, drift $0.1$, $K=\tau=5$, $\beta_{AV}\in\{0,0.5,1.0\}$, $\gamma_0\in\{-3.2,-2.0,-0.9\}$, $\gamma_L=1.0$ (core), $\gamma_V=0.5$, and $\varepsilon_k,L_0\sim N(0,1)$; the full parameterization and calibration are in Web Appendix~A (Web Table~S2).

\textbf{Correctness gate (base anchor).} A pre-specified base cell ($\beta_{AV}=0$, $\gamma_L=0$, minimal
switching, $\beta_A$ nonzero) validates the estimator limits
before any cell is reported: our implementation confirms that a Cox model adjusting for the time-updated
confounder $L_t$ recovers $\beta_A$ to Monte-Carlo error, and that the base-cell marginal-versus-conditional
separation---non-collapsibility \emph{plus} treatment-affected-$L$ mediation, since $\psi>0$ remains on when
$\gamma_L=0$---exceeds its Monte-Carlo standard error. Under the primary continuous-time Cox mechanism (DGP-A), every
correctly-specified estimator's mean absolute own-limit bias is at most $0.043$, within the pre-set $0.05$ bound (Web Appendix~C). Under the secondary discrete-time replication mechanism (DGP-B), where the continuous-time estimators are mildly misspecified by construction, the maximum is $0.059$ (Web Appendix~F). Because each own-limit truth carries a Monte-Carlo error of about $0.03$ (Web Appendix~E), the gate is a coarse consistency screen: it certifies own-limit bias only to about $0.05$--$0.06$, and we read the DGP-B maximum as indistinguishable from the bound rather than as clearing it.

\section{Simulation study}\label{sec:sim}

\textbf{Design.} Following the ADEMP framework (aims, data-generating mechanisms, estimands, methods, performance measures) \citep{morris2019}, we run a $31$-cell factorial that crosses
effect modification $\beta_{AV}$ (none / moderate / strong) and control-initiation rate $\kappa$ (approximately
$9\%/20\%/35\%$ per visit, set by calibrating the initiation intercept $\gamma_0$; Web Appendix~A) with sample size $n\in\{500,2000\}$, plus a base anchor, a strong-confounding slice,
and an unmeasured-frailty slice that adds an unmeasured \emph{prognostic} factor---entering the outcome only and
independent of treatment given the measured history---as a robustness check, on which every correctly coded
method stays own-limit consistent. Each cell runs
$1{,}000$ replicates with seeds shared across methods; the truth for each estimand is computed once per cell on a
$20{,}000$-subject pseudo-population, whose stability we verify (Web Appendix~A). Table~\ref{tab:simmap} maps each
experiment to the question it answers and where the result appears.

\begin{table}[htbp]\centering\small
\caption{\textbf{Roadmap of the simulation study.} Each experiment, the question it addresses and what it shows,
and the display item that reports it. The correctness gate is a prerequisite check (a consistent estimator cannot
fail it); the four numbered findings are the paper's contributions. $\kappa$ is the control-initiation rate, the
per-visit probability that a still-untreated control patient initiates treatment. ``\S'' denotes a main-text
section; ``Web'' items are in the Supplementary Material.}\label{tab:simmap}
\begin{tabular}{@{}>{\raggedright\arraybackslash}p{0.20\textwidth} >{\raggedright\arraybackslash}p{0.56\textwidth} >{\raggedright\arraybackslash}p{0.16\textwidth}@{}}
\toprule
\rowcolor{headGrey} Experiment & Question addressed, and what it shows & Location \\
\midrule
\multicolumn{3}{@{}l}{\emph{Prerequisite}}\\
Correctness gate & \emph{Is every estimator consistent for its own limit?} Yes; max mean $|$bias$|$ $0.043$ (DGP-A), $0.059$ (DGP-B). & \S\ref{sec:dgm}; Web~C \\
\addlinespace
\multicolumn{3}{@{}l}{\emph{Findings}}\\
Phantom bias & \emph{How much bias does a shared ``true HR'' fabricate?} $+0.33$ (ATT), $+0.15$ (conditional); the naive's $+0.76$ is total discrepancy, not phantom bias. & Table~\ref{tab:phantom} \\
Rank reversal & \emph{Does the best method flip with the target estimand?} Yes (bias-aligned); an MSE winner differs. & Table~\ref{tab:rankrev} \\
Variance result & \emph{Does the cheap sandwich cover?} For stacking yes; for matching no---bootstrap repairs it. & Table~\ref{tab:cover} \\
Model sensitivity & \emph{How much does omitting a measured confounder cost?} $0.45$--$0.50$ log-HR; coverage collapses. & \S\ref{sec:results}; Web~C \\
ITT vs.\ PP & \emph{How large is the ITT--PP gap, and when?} $-0.12$ RMST; grows in magnitude with $\kappa$. & \S\ref{sec:results}; Web~C \\
\bottomrule
\end{tabular}
\end{table}

\textbf{Estimands and performance measures.} Per cell we quantify the eight performance measures of
Table~\ref{tab:measures}: the own-limit correctness gate; \textbf{phantom bias} against the $s{=}0$ marginal truth;
the best method per target estimand (\textbf{rank reversal}) under two criteria, defined below; empirical standard
deviation and root mean-squared error (not printed here---this paper reports the derived variance- and
SE-ratios instead, with the full per-cell values in the code repository); confidence-interval coverage for the cluster-robust sandwich and the subject-level bootstrap, each with
its Monte-Carlo standard error (MCSE); a construction-failure rate (zero throughout); a \textbf{model-sensitivity}
comparison (each estimator under a correctly specified versus a deliberately misspecified adjustment or propensity
model, the induced bias being the sensitivity); and the \textbf{intention-to-treat versus per-protocol} contrast. Because subject-level bootstrap coverage costs an order
of magnitude more compute than the point metrics, it is run as a separate sub-study over a pre-chosen subset of
cells with $100$ outer replicates per cell, each bootstrapped with a pre-committed $B{=}100$ inner resamples; the
outer replicate count (not $B$) sets its Monte-Carlo error, which is sized to distinguish $90\%$ from $95\%$
coverage (Web Appendix~E).

\begin{table}[htbp]\centering\small
\caption{\textbf{Performance measures} (the ADEMP ``P''). Each is computed per cell and aggregated as in \S5. The tables in this paper report selected summaries; the complete per-cell values for every measure---including the empirical SD and RMSE, which this paper does not print---are provided as data files in the public code repository (Data Availability).}\label{tab:measures}
\begin{tabular}{@{}>{\raggedright\arraybackslash}p{3.3cm} >{\raggedright\arraybackslash}p{6.3cm} >{\raggedright\arraybackslash}p{3.2cm}@{}}
\toprule
\rowcolor{headGrey} Measure & Definition & Reported in \\
\midrule
Own-limit bias (\textbf{gate}) & mean signed bias against each estimator's own probability limit & \S4; Web App.~C \\
\textbf{Phantom bias} & bias against the shared $s{=}0$ marginal truth & Table~\ref{tab:phantom}; Web App.~C \\
\textbf{Rank reversal} & best method per target, bias-aligned and by MSE & Table~\ref{tab:rankrev}; Web App.~C \\
Empirical SD, RMSE & spread of $\hat\theta$ across the $1{,}000$ replicates & Code repository (per cell) \\
CI coverage $\pm$ MCSE & sandwich (full run) and bootstrap (sub-study) coverage of nominal $95\%$ & Table~\ref{tab:cover}; Fig.~\ref{fig:coverage} \\
Construction-failure rate & fraction of replicates with no estimate ($0$ throughout) & Web App.~E \\
\textbf{Model sensitivity} & misspecified minus correctly-specified own-limit bias & \S6; Web App.~C \\
\textbf{ITT vs.\ per-protocol} & marginal RMST contrast (ITT minus per-protocol) & \S6; Web App.~C \\
\bottomrule
\end{tabular}
\end{table}

\textbf{Two rankings for the reversal.} For each target estimand we rank the correctly-specified log-HR methods
on the \emph{full} pool (no per-target restriction), under two criteria. The \emph{bias-aligned} rank scores each
method by the distance of its mean estimate to the target's truth, $|\bar{\hat\theta}-\theta_{\text{tgt}}|$: the
winner is the method whose probability limit \emph{aligns} with the requested estimand, and the flip of that
winner across estimands is the estimand-aligned reversal. This is not the own-limit gate. Every method is scored against a
\emph{fixed} target, not its own moving truth. The \emph{MSE} rank scores by
$\sqrt{(\text{bias})^2+\text{SD}^2}$; here a precise estimator of a \emph{different} estimand can win, which is
the phantom-bias logic applied to method choice. Near-ties
(within twice the Monte-Carlo standard error of the winner) are recorded so a coincidental probability-limit
cluster is not read as a decisive flip.

\section{Results}\label{sec:results}

\textbf{Correctness gate.} For correctly-specified estimators, mean absolute own-limit bias is
$0.02$--$0.04$ on the log-HR / RMST scale, so the correctness gate passes and every effect below is
correctness-checked. The deliberately misspecified arms sit far higher, at $0.48$ and $0.50$ against the correct truth (a \emph{sensitivity} of $0.45$ and $0.50$ net of the correctly-specified bias; \S``Model sensitivity''), and are analysed separately.

\subsection{Phantom bias: a shared ``true HR'' fabricates bias}

\textbf{The magnitudes.} \emph{Scoring estimand-distinct methods against one marginal truth manufactures a third of a
log-cumulative-hazard ratio of bias that is entirely an estimand mismatch.} Scored against a single shared
$s{=}0$ marginal truth on the log-cumulative-hazard-ratio (log-CHR) scale (Table~\ref{tab:phantom}), the two ATT
matching methods show $+0.33$ and $+0.32$ and the conditional method $+0.15$, because each targets a parameter that is
numerically distinct from the marginal benchmark, not because any is inconsistent. The naive time-dependent Cox sits
$+0.76$ from the same reference, but that figure is a \emph{total discrepancy}: its limit is associational, so the
gap compounds the estimand mismatch with uncontrolled treatment-affected confounding and is not an instance of the
Corollary.

\textbf{The marginal rows are not exempt.} The marginal methods are, by
construction, near zero on their own scale; we place them on the fixed-horizon log-cumulative-hazard-ratio scale so
the comparison is like-for-like. Their small residual ($+0.12$ for the marginal structural model, $+0.20$ for
the standardized Sequential Cox) has the same origin as the ATT and conditional rows. It is an estimand gap, not an
artifact of how the effect is summarized. All three quantities are the \emph{same} fixed-horizon log-CHR functional;
they differ because each is evaluated on a different constructed population: the inverse-probability-weighted full
cohort for the marginal structural model, the origin-pooled trial-eligible cohort for the standardized Sequential
Cox, and the coupled always-versus-never pseudo-population for the shared reference. That population difference
accounts for the bulk of each marginal residual: about $0.08$ of the marginal structural model's $+0.12$ and $0.17$ of the standardized Sequential Cox's $+0.20$, with the marginal non-proportionality quantified in Web Appendix~C (about $0.03$) and a small interaction making up the remainder (see ``What is being differenced, exactly'' below). A published
side-by-side that compared these ATT methods to a marginal benchmark would thus report roughly a third of a
log-CHR of ``bias'' that is an artifact of the comparison, not a property of the estimators.

\textbf{What survives a change of scale.} Because a risk
difference, survival difference, or restricted-mean-survival-time difference is \emph{collapsible}, its marginal contrast equals an appropriately standardized average of the conditional contrasts, with no non-collapsibility gap (and coincides with the conditional contrast when the effect is homogeneous); such a summary therefore carries no non-collapsibility
component of phantom bias, and the pathology quantified here is specific to the non-collapsible hazard ratio. The
remaining sources---treatment-affected-confounder mediation, ATT selection, and ITT dilution---are genuine
estimand differences that persist on any scale.

\textbf{What is being differenced, exactly.} The shared reference must be a \emph{functional} of the
counterfactual survival curves---here $\theta^{\mathrm{marg\text{-}s0}}=\log\{-\log S^1(\tau)\}-\log\{-\log
S^0(\tau)\}$, estimated by Kaplan--Meier on the coupled counterfactual pseudo-population---because every method is
scored against it, and a reference defined by a fitted working model would make the ``truth'' depend on which
model happened to be fitted. The estimator \emph{limits} are a different matter. Sequential stratification,
time-dependent propensity-score matching, and the conditional analysis define their targets \emph{through the
models they fit}, so their limits are (stratified-)Cox partial-likelihood projections rather than functionals of
$S^1,S^0$; the phantom bias reported for those rows therefore carries a summary-functional component alongside the
target-population component. This is the paper's central point: an estimator
whose target is defined by a working model \emph{has} a model-defined estimand, and no single ``true hazard
ratio'' can score it. Web Appendix~C reports, for each mechanism, the range and mean across cells of the gap between
the functional reference and the probability limit of a marginal Cox fit to the same counterfactuals; that gap is
negative in every cell; it is the marginal non-proportionality this mechanism induces, and it is why the reference
is not itself a Cox coefficient.

\begin{table}[htbp]\centering\small
\caption{\textbf{Phantom bias (31-cell run).} Mean estimate minus the $s{=}0$ marginal truth
$\theta^{\mathrm{marg\text{-}s0}}=\log\{-\log S^1(\tau)\}-\log\{-\log S^0(\tau)\}$, a Kaplan--Meier plug-in
functional on the log-cumulative-hazard-ratio scale; all rows are differenced against that same reference.
Phantom bias is the decision-relevant bias a shared-``true-HR'' comparison fabricates; it is
\emph{not} a consistency failure (each causal method's own-limit bias is $\le0.043$ (DGP-A) and $\le0.059$ (DGP-B, within Monte-Carlo error of the $0.05$ gate); Table~\ref{tab:simmap}). The ATT,
conditional and marginal rows show the mismatch a marginal benchmark would misattribute as method bias. The naive
row is separated below the rule and shaded: its probability limit is \emph{associational}, so its entry is not a
phantom bias but a \emph{total discrepancy}---the estimand gap compounded with uncontrolled treatment-affected
confounding---and it is not an instance of the Corollary.}\label{tab:phantom}
\begin{tabular}{@{}>{\raggedright\arraybackslash}p{7.6cm}c@{}}
\toprule
\rowcolor{headGrey} Method & Discrepancy vs.\ $s{=}0$ marginal truth (log-CHR) \\
\midrule
Sequential stratification, ATT & $+0.33$ \\
TD-PS matching, ATT & $+0.32$ \\
Sequential Cox, conditional & $+0.15$ \\
\addlinespace
Standardized Sequential Cox (marginal) & $+0.20$ \\
IPTW marginal structural model & $+0.12$ \\
\midrule
\multicolumn{2}{@{}l}{\emph{Total discrepancy (associational own-limit; not covered by the Corollary)}}\\
\cellcolor{warnTint}Naive time-dependent Cox & \cellcolor{warnTint}$+0.76$ \\
\bottomrule
\end{tabular}
\end{table}

\subsection{Rank reversal: the best method flips with the target estimand}

\begin{figure}[t]\centering
\includegraphics[width=\linewidth,alt={A two-column tile grid over six target-estimand rows. The left
(bias-aligned) column changes colour down the rows, showing the winning method flips with the estimand; the right
(mean-squared-error) column repeats the low-variance ITT estimator on three of six rows, including two it does not target.}]{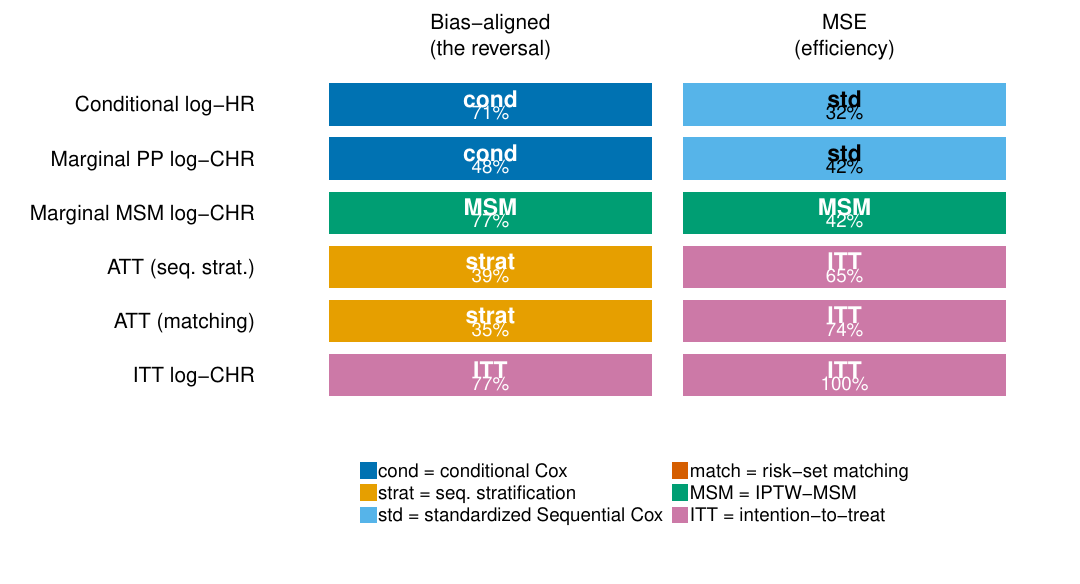}
\caption{\textbf{Rank reversal.} For each target estimand (rows), the method winning the most of the $31$ cells
under the \emph{bias-aligned} criterion (left, which estimator's probability limit is closest to the target) and the \emph{mean-squared-error} criterion (right, which adds efficiency), with its win
percentage. Under bias alignment the winner flips with the estimand---conditional and per-protocol-marginal targets to the conditional Cox (cond) and the marginal-MSM target to the IPTW-MSM (MSM), the ATT targets to sequential stratification
(strat), and the ITT target to the ITT estimator---whereas under MSE the low-variance ITT estimator dominates
several targets it does not target.}
\label{fig:rankrev}
\end{figure}

\emph{The bias-aligned winner changes with the target estimand.} Table~\ref{tab:rankrev} and
Figure~\ref{fig:rankrev} report, for each target estimand, the method that wins most of the $31$ cells under the
two criteria of Section~\ref{sec:sim}. On the \textbf{bias-aligned} criterion---which method's probability limit is closest to the
requested estimand---the winner flips across the estimand space: the conditional target is won by the conditional
Cox ($71\%$ of cells), the marginal-structural-model target by the IPTW-MSM ($77\%$), the per-protocol marginal by
the conditional and standardized estimators, the two ATT targets by the matching estimators (sequential
stratification leading, $39\%$ and $35\%$, with risk-set matching a close second), and the intention-to-treat target by the
ITT estimator ($77\%$). In this mechanism the ATT and ITT probability limits nearly coincide ($\approx-0.44$ to $-0.45$)
while the conditional/marginal limits cluster at $\approx-0.60$ to $-0.66$, so the reversal is sharp \emph{between}
these two clusters (a $0.15$--$0.22$ log-HR gap between clusters, far exceeding Monte-Carlo error) and the fine ordering
\emph{within} the coincident cluster is a near-tie; that the ATT and ITT estimands collapse
under modest switching is a finding in its own right.

\textbf{MSE: efficiency is not alignment.} On the \textbf{MSE} criterion the picture differs: a low-variance
estimator wins targets it does not aim at. The intention-to-treat estimator, which pools all trials without
per-protocol censoring and so has the smallest empirical standard error, takes the two ATT targets ($65\%$ and
$74\%$); the standardized Sequential Cox takes the conditional target ($32\%$) it does not aim at (a precise
estimate of a nearby estimand beating a noisier on-target one), while on the per-protocol-marginal
target ($42\%$) it is the on-target winner. This is the phantom-bias logic applied to method choice: optimizing ``distance to some hazard
ratio'' rewards precision over estimand alignment, which is exactly why the estimand must be named first. Web
Appendix~A (Theorem~1) states both halves: no estimator is bias-optimal across distinct estimand limits, and an
off-target estimator wins on MSE when its squared bias to the target falls below the variance it saves (for an
unbiased on-target estimator, $\Delta^2<V_{\mathrm{on}}-V_{\mathrm{off}}$, with $\Delta$ the probability-limit
distance to the target and $V_{\mathrm{on}}/V_{\mathrm{off}}$ the on-/off-target variances), an accounting that correctly predicts which of the two---sequential stratification (on-target) or the ITT estimator (off-target)---has the lower RMSE to the ATT truth in $30$ of $31$ cells, driven by the matching estimators' $\approx3.4$--$4.5\times$ larger variance than the
trial-pooled ITT estimator.

\begin{table}[htbp]\centering\small
\caption{\textbf{Rank reversal (31-cell run).} For each target estimand, the method winning the most of $31$
cells under two criteria, with its win count out of the $31$ cells (win fraction in parentheses). \emph{Bias-aligned}
(shaded): the method whose probability limit is closest to the target; the winner flips
across estimands. \emph{MSE}: lowest root-mean-squared error to the target, which adds efficiency; on three of the
six targets the winner is a low-variance estimator aimed elsewhere: the ITT estimator on the two ATT targets and
the standardized Sequential Cox on the conditional target; on the remaining three the on-target estimator also
wins on MSE, an efficiency effect (a precise estimate of a nearby
estimand can beat a noisier on-target one). The conditional/marginal targets (limits $\approx-0.60$ to $-0.66$)
and the ATT/ITT targets (limits $\approx-0.44$ to $-0.45$) form two clusters; the reversal is between clusters, and
within-cluster orderings are near-ties.}\label{tab:rankrev}
\begin{tabular}{@{}>{\raggedright\arraybackslash}p{4.0cm} l l@{}}
\toprule
\rowcolor{headGrey} Target estimand & Bias-aligned winner (the reversal) & MSE winner (efficiency) \\
\midrule
Conditional log-HR & \cellcolor{cfTint}conditional Cox, $22/31$ ($71\%$) & standardized Seq.\ Cox, $10/31$ ($32\%$) \\
Marginal log-CHR (PP) & \cellcolor{cfTint}conditional Cox, $15/31$ ($48\%$)$^\dagger$ & standardized Seq.\ Cox, $13/31$ ($42\%$) \\
Marginal log-CHR (MSM) & \cellcolor{cfTint}IPTW-MSM, $24/31$ ($77\%$) & IPTW-MSM, $13/31$ ($42\%$) \\
ATT, seq.\ stratification & \cellcolor{cfTint}seq.\ stratification, $12/31$ ($39\%$)$^\ddagger$ & ITT, $20/31$ ($65\%$) \\
ATT, TD-PS matching & \cellcolor{cfTint}seq.\ stratification, $11/31$ ($35\%$)$^\ddagger$ & ITT, $23/31$ ($74\%$) \\
Intention-to-treat log-CHR & \cellcolor{cfTint}ITT, $24/31$ ($77\%$) & ITT, $31/31$ ($100\%$) \\
\bottomrule
\end{tabular}
\\[3pt]{\footnotesize $^\dagger$ the standardized Sequential Cox is the close runner-up ($11/31$). $^\ddagger$ TD-PS matching
is the close runner-up ($10/31$); the two matching estimators together win the ATT targets.}
\end{table}

\subsection{A cross-family variance result: the sandwich under-covers matching, the bootstrap repairs it}

\begin{figure}[t]\centering
\includegraphics[width=\linewidth,alt={Dot-and-error-bar plot of coverage for six estimators. Cluster-robust
sandwich coverage is about 0.90 for the trial-stacking conditional estimator but drops to about 0.77 and 0.82 for
the two matching (ATT) estimators; subject-level bootstrap coverage is 0.91 to 0.96 for every estimator.}]{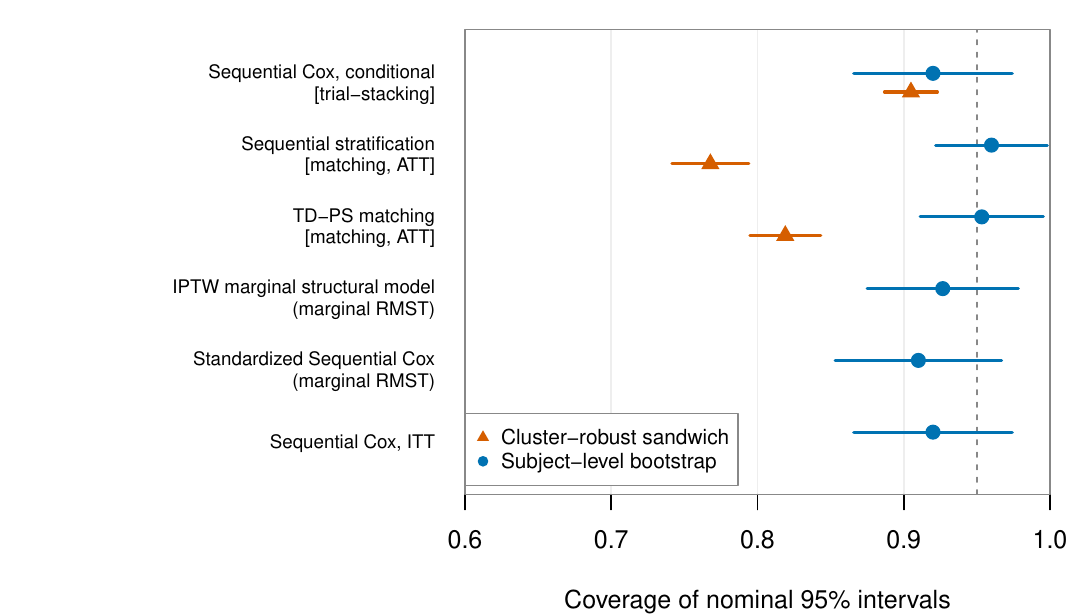}
\caption{\textbf{The cross-family variance result.} Coverage of nominal $95\%$ intervals for each estimator under the
cluster-robust sandwich (red triangles, full $31$-cell run) and the subject-level bootstrap (blue circles,
pre-specified $n{=}500$ sub-study), with $\pm2$ Monte-Carlo standard errors on the same per-cell basis as
Table~\ref{tab:cover} (the mean per-cell binomial MCSE, the precision of a typical cell, not of the pooled mean,
so these bars are the conservative choice); the dashed line marks $0.95$ (the two
interval methods are defined in Table~\ref{tab:cover}). The sandwich is closer to nominal for the trial-stacking
conditional estimator but under-covers the two matching (ATT)
estimators; the subject-level bootstrap repairs them and brings the marginal / RMST estimators (no closed-form
sandwich) to near-nominal ($\approx0.91$--$0.94$). In the cells studied, the cluster-robust sandwich suffices for the trial-stacking estimators while the matching estimators need the subject-level bootstrap.}
\label{fig:coverage}
\end{figure}

\emph{The cheap cluster-robust sandwich is closer to nominal for trial-stacking estimators than for matching
estimators, for which the subject-level bootstrap restored near-nominal coverage in the cells we examined.} The cluster-robust sandwich is closer to
nominal for the trial-stacking conditional estimator ($0.90$) than for the matching estimators
($0.77$ for sequential stratification, $0.82$ for risk-set matching; Table~\ref{tab:cover} and
Figure~\ref{fig:coverage}), though still below the $0.95$ target, even though the
matching estimators reuse subjects \emph{less} than the trial-stacking one (mean reuse $1.3$--$1.9$ versus $2.8$);
the failure is therefore not a reuse phenomenon but the treatment of estimated matched sets as fixed
\citep{abadie2006}. Web Appendix~B gives the mechanism as a heuristic: the matching variance has a within-set
component the sandwich captures and an \emph{assignment} component---from the matched sets being data-estimated
(the matching uncertainty of \citealp{abadie2006})---that it misses, zero only when the strata are design-fixed, so
the sandwich understates the matching standard error.

\textbf{What the bootstrap evidence supports.} The sandwich-SE/SD ratios ($0.63$--$0.69$ for the
estimated-matched-set estimators versus $0.90$ for the design-fixed trial origins) evidence it.

The subject-level bootstrap repairs both: it resamples whole subjects, so it respects the reuse clustering, and---unlike the sandwich---re-forms the matched sets on each resample, propagating the set-estimation uncertainty the sandwich conditions away. Coverage returns to $0.96$ and $0.95$ for the matching estimators and to $0.91$--$0.94$ for the marginal/RMST estimators, which have no closed-form sandwich. We report that repair as an \emph{empirical} result: a full
asymptotic justification of the bootstrap for a data-adaptive matching functional is not attempted here, and
\citet{abadie2008} show the pair bootstrap can fail for fixed-$M$ nearest-neighbour matching, a different
construction, but a reason to treat the repair as demonstrated rather than proved (Web Appendix~B). In the cells we examined the pattern is
consistent: the cluster-robust sandwich for trial-stacking estimators and the subject-level bootstrap for
matching estimators---a practical guide, not a claim we establish beyond those cells.

\begin{table}[htbp]\centering\small
\caption{\textbf{Coverage of nominal $95\%$ intervals.} Cluster-robust sandwich (full $31$-cell run) versus
subject-level bootstrap (pre-specified sub-study at sample size $n{=}500$, $100$ outer replicates each with $B{=}100$ inner resamples; the sub-study is cells $4$, $6$, $9$ of Web Table~S2, with $(\beta_{AV},\gamma_0)=(0,{-}2.0),(1,{-}2.0),(1,{-}0.9)$). Each entry is the unweighted mean of the
per-cell coverages, and the figure in parentheses is the mean of the per-cell binomial Monte-Carlo standard errors
$\sqrt{p(1-p)/n_{\mathrm{rep}}}$ ($n_{\mathrm{rep}}=1{,}000$ for the full-run sandwich, $100$ outer replicates for the bootstrap sub-study, which is why the bootstrap parentheses are larger), that is, the precision of a \emph{typical cell}, not of the pooled mean, which
is smaller by roughly $\sqrt{\text{number of cells}}$. Coverage varies appreciably across cells (sandwich ranges
$0.78$--$0.95$ trial-stacking, $0.52$--$0.86$ sequential stratification, $0.74$--$0.87$ matching), so the
cell-to-cell spread exceeds the Monte-Carlo error; the cross-family gap is nonetheless many standard errors wide
on any basis. Restricting the sandwich to the same three sub-study cells gives $0.91$, $0.78$, $0.82$ (conditional,
sequential stratification, TD-PS matching), essentially the all-cell means, so the sandwich and bootstrap are
compared on comparable ground.
The \emph{cluster-robust sandwich} is the model-based robust variance clustered on the subject (each
subject may enter several trials or matched sets), computed with the estimated trials/matched sets held fixed; the
\emph{subject-level bootstrap} resamples whole subjects and re-forms those trials/matched sets on each resample,
additionally capturing the uncertainty in the estimated matched sets. The sandwich covers the trial-stacking
conditional estimator but under-covers the matching
estimators (shaded); the bootstrap repairs them. ``---'' denotes no closed-form sandwich (marginal/RMST
estimators rely on the bootstrap). Reuse (last column) is the mean number of trials/sets a subject enters, reported descriptively only (for the matching estimators averaged over matched controls, for trial-stacking over all enrolled subjects, so the two are not on a common denominator). On a common basis, the sandwich-SE/SD ratios---$0.63$--$0.69$ for the matching estimators versus $0.90$ for trial-stacking---locate the under-coverage in the treatment of estimated matched sets as fixed rather than in reuse.}\label{tab:cover}
\begin{tabular}{@{}>{\raggedright\arraybackslash}p{6.6cm} c c c@{}}
\toprule
\rowcolor{headGrey} Method & Sandwich & Bootstrap & Reuse \\
\midrule
Sequential Cox, conditional, trial-stacking & $0.90$ {\scriptsize(0.01)} & $0.92$ {\scriptsize(0.03)} & $2.8$ \\
Sequential stratification, ATT & \cellcolor{warnTint}$0.77$ {\scriptsize(0.01)} & $0.96$ {\scriptsize(0.02)} & $1.9$ \\
TD-PS matching, ATT & \cellcolor{warnTint}$0.82$ {\scriptsize(0.01)} & $0.95$ {\scriptsize(0.02)} & $1.3$ \\
\addlinespace
IPTW marginal structural model & --- & $0.93$ {\scriptsize(0.02)} & $1.0$ \\
Standardized Sequential Cox (marginal RMST) & --- & $0.91$ {\scriptsize(0.03)} & $2.8$ \\
Sequential Cox, ITT & --- & $0.94$ {\scriptsize(0.03)} & $2.8$ \\
\bottomrule
\end{tabular}
\end{table}

\textbf{Model sensitivity.} \emph{The methods are highly sensitive to correct specification of the
adjustment/propensity model, an axis the estimand-alignment view does not capture.} Refitting the conditional
sequential-Cox estimator while omitting the baseline confounder from the adjustment set, and the propensity-score
matcher while omitting it from the propensity model, induces a bias of $0.45$ and $0.50$ on the log-HR scale
(against a correctly-specified bias near zero), with coverage collapsing to $0.19$ and $0.12$ under the sandwich,
because the misspecified model no longer targets the intended limit. The full arm is in Web Appendix~C.

\textbf{Intention-to-treat versus per-protocol.} \emph{The ITT--PP gap is a quantity, not a labelling remark.}
The ITT estimator (no artificial censoring of control initiators) and the per-protocol estimator coincide when
control initiation is rare and diverge as it becomes common; over the run the marginal RMST gap is $-0.12$
($0.371$ ITT versus $0.495$ PP) and widens in magnitude with the control-initiation rate $\kappa$, tracing the
gap directly rather than leaving it implicit.

\textbf{External validity.} \emph{Three of the four findings replicate on a structurally different
mechanism---two directly and the rank reversal in attenuated form---while the model-sensitivity finding does not carry over
in magnitude.} Replicating the study on a discrete-time pooled-logistic hazard with a binary time-varying
confounder---which renders the continuous-time Cox and matching estimators mildly misspecified---reproduces the
phantom bias (correctly signed, though milder) and the cross-family variance result directly: the sandwich still
under-covers the matching estimators while the bootstrap repairs them (the IPTW marginal structural model's
phantom-bias offset reverses sign across mechanisms, so we claim replication of the phenomenon, not of every
per-method sign). The rank reversal is present but weaker: the bias-aligned winner still changes with the target,
but over a narrower range (three distinct winners across the six targets, against four under the primary
mechanism), as the milder calibration places the estimand limits closer together. And the model-sensitivity and
intention-to-treat-versus-per-protocol finding does not carry over in magnitude: DGP-B holds confounding
deliberately mild, so the misspecification sensitivity shrinks to $0.014$ and $0.018$ (against $0.45$ and $0.50$)
and the ITT--PP marginal RMST gap to $-0.030$ (against $-0.124$), consistent with the paper's thesis that the
qualitative patterns are calibration-independent while their magnitudes are not. The findings are therefore not
artifacts of the continuous-time Cox construction (the four-finding comparison is Web Table~S12; Web Appendix~F).

\section{Application: Stanford heart transplant}\label{sec:applied}

\textbf{A regime in which the methods do not diverge---and a negative control showing what mis-alignment would have
manufactured.} We apply the four families to the Stanford heart transplant data \citep{crowley1977} ($103$
patients, $69$ transplants, $75$ deaths), with transplant as a once-on time-dependent treatment, age (standardized) as a baseline
confounder, and prior bypass surgery as a baseline prognostic covariate; follow-up is binned into $30$-day visits to
match the estimators' structure (the median wait to transplant is $26$ days). A subject counts as treated only
from the first visit interval \emph{beginning at or after} transplant.

\emph{No interval excludes the null, and the estimates are imprecise.} The estimates cluster near the null and every interval covers it
(subject-level bootstrap $95\%$ intervals, $B{=}400$):
conditional hazard ratio $0.94$ $[0.49,2.09]$, average-treatment-effect-on-the-treated hazard ratios $0.67$
$[0.15,2.25]$ (sequential stratification) and $0.81$ $[0.32,1.90]$ (risk-set matching), and marginal
restricted-mean-survival-time differences of $+0.08$ $[-0.91,0.96]$ and $0.00$ $[-4.68,3.14]$ years.

\textbf{Interpretation.} The
families do not diverge here, but with $103$ patients, hazard-ratio point estimates spanning $0.67$--$0.94$, and
every interval covering the null, this near-null agreement is consistent with explanations these data cannot
separate: a genuine null, limited power, and the absence of a strong measured time-dependent confounder. Near the
null the estimand gaps our simulation exhibits are expected to be small here---the non-collapsibility component shrinks with the effect size, and the measured time-dependent confounding in these data is modest---so the aligned analysis
\emph{illustrates}, rather than tests, the convergence the simulation shows absent strong time-dependent
confounding; it is not evidence of a transplant effect.

\textbf{Causal caveat.} Read causally, these estimates further rest on strong,
unverifiable assumptions in these data: no unmeasured confounding of transplantation given age, consistency, and
positivity.

\emph{Negative control.} Coding treatment the way a wide bin and a mid-interval transplant invite---crediting the
transplant from the \emph{start} of the interval containing it, at a $180$-day bin---backdates $67$ of the $69$
recipients to time zero and manufactures exactly the strong benefit the aligned analysis does not find:
conditional hazard ratio $0.24$, ATT $0.15$ and $0.18$, and restricted-mean differences of
$+1.42$ and $+1.12$ years. This is immortal-time bias, the pathology named in Section~\ref{sec:intro}: the
recipients' entire pre-transplant survival is credited to the treated arm. We include it because it
illustrates the same problem as the rest of the paper: a benchmark number can be an artifact of a
modelling choice (there, which estimand is scored; here, when treatment is deemed to start) rather than a property
of any estimator. Both analyses are reproducible from the accompanying code
(reproduction commands in the reproducibility guide). Web Appendix~D gives the binning, eligibility, and per-method estimates with subject-level bootstrap intervals, plus a bin-width sensitivity note.

\section{Discussion}\label{sec:discussion}

\textbf{Summary.} Longitudinal-matching and target-trial-emulation methods are not interchangeable estimators of
one quantity; they answer different causal questions, and a benchmark that ignores this manufactures bias. The
own-limit comparison is a correctness gate, not a result; the results are the phantom-bias magnitude, the
estimand-dependent rank reversal (with efficiency separated from alignment), the cross-family variance result, and
the model-sensitivity and ITT-versus-PP arms.

\textbf{Practical implications.} The method follows from the causal question, not the reverse;
Table~\ref{tab:recommend} is a by-estimand guide. Where no single
recommendation is possible we say so: no method is best across estimands, and the adjustment / propensity
model must be correct for any of them (\S\ref{sec:results}).

\begin{table}[htbp]\centering\small
\caption{\textbf{Choosing a method: a practical guide by target estimand.} There is no single ``best'' method: the
choice follows from the causal question being asked. For \emph{every} row, the adjustment / propensity model must
be correctly specified and there must be no unmeasured confounding of initiation given the measured history
(omitting a confounder biases the estimate and collapses coverage; \S\ref{sec:results}), and
selecting a method by lowest mean-squared error to ``some hazard ratio'' is \emph{not} a substitute for naming the
estimand (a precise estimator of a nearby estimand can win on MSE while answering a different
question).}\label{tab:recommend}
{\setlength{\tabcolsep}{4pt}%
\begin{tabular}{@{}>{\raggedright\arraybackslash}p{3.9cm}>{\raggedright\arraybackslash}p{3.2cm}>{\raggedright\arraybackslash}p{2.8cm}>{\raggedright\arraybackslash}p{4.0cm}@{}}
\toprule
\rowcolor{headGrey} Your causal question (target estimand) & Recommended estimator & Confidence interval & Key caveat \\
\midrule
Covariate-\emph{conditional} hazard ratio (adjusted, within-stratum effect) & Sequential Cox, conditional & Cluster-robust sandwich ($0.90$; mildly anti-conservative) or bootstrap & Do not benchmark it against a marginal or ATT value---the gap is non-collapsibility, not bias \\
\addlinespace
\emph{Marginal} population-average effect (survival difference, RMST, or cumulative-HR) & Standardized Sequential Cox, or IPTW-MSM & Subject-level bootstrap (no closed-form sandwich) & Differs from the conditional HR even absent confounding; a risk/RMST difference is collapsible, so it carries no non-collapsibility phantom bias (mediation, selection, and dilution gaps persist on any scale) \\
\addlinespace
Effect \emph{among the treated} (ATT)---those who actually initiate & Sequential stratification, or TD-PS matching & Subject-level bootstrap (near-nominal in the cells studied) & The cluster-robust sandwich under-covers here ($\approx0.77$--$0.82$); do not rely on it \\
\addlinespace
\emph{As-assigned} (intention-to-treat) effect of the starting strategy & Sequential Cox, ITT & Subject-level bootstrap & Diluted toward the null relative to per-protocol as switching ($\kappa$) rises \\
\bottomrule
\end{tabular}}
\end{table}

\textbf{Limitations and future work.} We rely on a Cox-native mechanism (additive-hazard inference is treated
elsewhere), a single time-varying confounder, and the discretization required to apply the estimators to
continuously observed data; the external-validity replication above shows the findings are not artifacts of the
continuous-time Cox construction (Web Appendix~F). In the primary continuous-time mechanism the ATT and ITT
probability limits nearly
coincide; a heavier-non-adherence regime that separates them---where the matching estimators would reclaim the ATT
target on both criteria---is flagged as a robustness extension.
We address all three of Thomas et al.'s named gaps: relative efficiency (the cross-family variance-ratio,
$3.4$--$4.5\times$), variance estimation (the sandwich-SE/SD ratios, $0.63$--$0.69$ for the matching estimators
versus $0.90$ for trial-stacking), both in \S\ref{sec:results}, and model
sensitivity. Extensions include dependent censoring, competing risks, and standardization-population variability.

\textbf{Conclusion.} On a non-collapsible survival mechanism, the question ``which method is best?'' is
ill-posed; the well-posed question is ``best for which estimand, and with which variance estimator?''---and this
benchmark makes it answerable.

\backmatter
\section*{Data Availability Statement}
Code and data are available at \url{https://github.com/ehsanx/lmtte}. The Stanford heart transplant data are distributed with the
R \texttt{survival} package \citep{crowley1977}.
\vspace*{-8pt}

\section*{Acknowledgements}
This research was supported in part through computational resources from Advanced Research Computing at the
University of British Columbia. During this work the author used AI-based tools to assist with analysis and
simulation code, text editing, and checking derivations; the author verified all outputs and takes full
responsibility for the content.
\vspace*{-8pt}

\section*{Funding}
This research received no specific grant from any funding agency in the public, commercial, or not-for-profit
sectors.
\vspace*{-8pt}

\section*{Conflict of Interest}
The author declares no potential conflict of interest.
\vspace*{-8pt}

\section*{Supplementary Material}
Web Appendices~A--F, Web Tables~S1--S13, and Web Figures~S1--S2 accompany this paper: (A) the data-generating
mechanism, known-truth construction, and the phantom-bias derivation; (B) estimator and variance definitions;
(C) the pooled simulation results (with the full per-cell values in the code repository), the diagnostic comparing the $s{=}0$ marginal functional
reference with the probability limit of a marginal Cox fit to the same counterfactuals (the marginal
proportional-hazards gap), and the model-sensitivity and ITT-versus-PP arms; (D) the Stanford
heart-transplant analysis; (E) the compute-time and robustness benchmark; and (F) an external-validity
test of all four findings---three replicate, while the model-sensitivity finding does not carry over in magnitude---under a second, structurally different data-generating mechanism.
\vspace*{-8pt}

\bibliography{references}
\label{lastpage}
\end{document}